\documentstyle{article}

\def\be{\begin{eqnarray}}
\def\ee{\end{eqnarray}}

\hoffset= - 1.0in         

\begin{document}

\hfill ITEP/TH-24/08

\bigskip \bigskip

\centerline{\Large{
NSR measures on hyperelliptic locus and
non-renormalization of 1,2,3-point functions
}}

\bigskip

\centerline{A.Morozov}

\bigskip

\centerline{{\it ITEP, Moscow, Russia}}

\bigskip

\centerline{ABSTRACT}

\bigskip

{\footnotesize
We demonstrate (under a modest assumption)
that the sums over spin-structures
of the simplest combinations of fermionic correlators
(Szego kernels) and DHP/CDG/Grushevsky NSR measures
vanish at least on the hyperelliptic loci in the
moduli space of Riemann surfaces -- despite the
violation of the $\theta_e^4$ hypothesis at $g>2$.
This provides an additional important support to
validity of these measures and is also a step towards
a proof of the non-renormalization theorems in the NSR
approach.
}

\bigskip

\bigskip

{\bf 1. Introduction.}
According to \cite{Mumm}-\cite{RScalc}, the Feynman diagram technique
in perturbative string theory in critical dimensions
is formulated in terms of holomorphic measures on the
moduli space ${\cal M}_g$ of Riemann surfaces (complex curves)
of genus $g$, where $g$ is the number of string loops:
somewhat symbolically string amplitudes
in the three simplest cases of $26d$ bosonic, $10d$ super- and
heterotic
models are given by integrals over ${\cal M}_g$ with the measures
\be
\frac{1}{{\det}^{13} {\rm Im}(T)}\Big|d\mu_{bos}\Big|^2\!\!, \ \ \ \
\frac{1}{{\det}^{5} {\rm Im}(T)}\left|\sum_e \ldots d\mu_e\right|^2
\!\!\!, \ \ \ \
\frac{1}{{\det}^{5} {\rm Im}(T)}\
\overline{d\mu_{het}} \left(\sum_e \ldots d\mu_e\right)
\ee
and appropriate vertex-operator insertions.
The basic Mumford measure
$d\mu_{bos} = \frac{\det\bar\partial_2}{\det^{13}\bar\partial_0}$
is well known, though in a somewhat transcendental form,
which, however simplifies for low genera $g\leq 4$ \cite{Mumm}
and on hyperelliptic loci ${\cal H}_g\subset {\cal M}_g$ \cite{LM}
of complex codimension $g-2$.
The measures for two basic heterotic models are given by \cite{Mumm}
\be
d\mu_{het}^{E_8\times E_8} = \xi_4^2\, d\mu_{bos}, \ \ \ \
d\mu_{het}^{SO(32)} = \xi_8\, d\mu_{bos}
\ee
which actually coincide (at least are believed to coincide)
on ${\cal M}_g$. Further compactifications to lower critical
dimensions involve additional modular-form factors in the measures,
see, for example, \cite{MO}. The same measures, restricted to the
moduli subspace of {\it doubles}, describe open and non-oriented
strings \cite{MR}.
The NSR measures $d\mu_e$, depending on the boundary conditions
for $2d$ fermions through the spin-structure (half-integer
theta-characteristic) $e$, remained a puzzle until a breakthrough
consideration of E.D'Hoker and D.Phong \cite{DHP1,DHP6}, and now they
are known (conjectured) \cite{CDG,Gru} in a very explicit form,
see also \cite{DHPoth,SuMe}.
The sum over $e$ describes GSO projection, which eliminates
tachyon excitation from the spectrum and makes the theories
supersymmetric in 10 dimensions (where they are alternatively
described by Green-Schwarz formalism \cite{GS}).
In the absence of vertex operators $\sum_e d\mu_e = 0$.
See \cite{SuMe} for notations, explanations, references and
further details.

The goal of the present paper is to consider restrictions of
DHP-CDG-G NSR measures on hyperelliptic loci
${\cal H}_g\subset{\cal M}_g$,
where they become pure algebraic quantities, made from
ramification points. This is a natural and standard
step in exploration of any theta-function-based formulas in
string theory, and it sheds light on the otherwise obscure
properties of the measures. Also properties and implications
of GSO projection become transparent in these restrictions.
All this is especially important because for $g>2$ the
DHP-CDG-G conjectures has some {\it a priori} unexpected properties
(most important, $d\mu_e$ does not contain a $\theta_e^4(0)$ factor,
as many people thought it would, and it is interesting to see
how the non-renormalization theorems \cite{Mart} are consistent with
this). Of course, hyperelliptic considerations are not
conclusive for $g>3$ (at $g=3$ they describe what happens
at codimension-one subspace in ${\cal M}_3$, and this is often
enough to draw far-going conclusions), still they are very
instructive.

\bigskip

{\bf 2. DHP/CDG/G conjectures and non-renormalization theorems.}
If NSR correlators are defined from supermoduli integration,
there are numerous different contributions \cite{DHP6}.
Most of them vanish after GSO summation over boundary
conditions (theta-characteristics) due to Riemann
identities.
This cancelation becomes transparent for special
choice of odd moduli ("unitary gauge"), largely for the
reasons which were outlined long ago in \cite{80s}.
There is however a non-trivial part, associated with
the DHP-NSR measure
\be
d\mu_e = \Xi_e \cdot d\mu_{bos}
\label{dmuvsBos}
\ee
where  $e$  is the even theta-characteristic,
$d\mu_{bos}$ is the Mumford measure \cite{Mumm}
and $\Xi_e$ is the weight-eight modular form,
discovered in \cite{DHP1,CDG,Gru}, which in Grushevsky
basis \cite{Gru} can be written as \cite{SuMe}
\be
\Xi_e =  \sum_{p\geq 0} (-)^p\kappa_p G^{(\!p\,)}_e, \ \ \ \ \ \ \ \
\kappa_p = \prod_{i=1}^p \frac{1}{2^i-1}, \ \ \ \ \kappa_0=1
\label{dmuGru}
\ee
The sum is actually up to $p=g$ where $g$ is the genus
(the number of string loops), because Grushevsky's
forms $G^{(\!p\,)} = 0$ for $p>g$.
If $\Xi_e$ in (\ref{dmuGru}) is multiplied by $2^{-g}$
the measures (\ref{dmuvsBos}) are properly factorized at
the boundaries of moduli space ${\cal M}_g$,
where genus-$g$ Riemann surface degenerates
into low-genera surfaces.

Because of the modular ambiguity of contributions,
annihilated in GSO projection by Riemann identities
\cite{ambig}, one can optimistically {\it assume} that they
do {\it not} contribute to any correlators at all,
so that for any observable $A$ in superstring theory
\be
<A> \ = \sum_e \int_{{\cal M}_g} A_e\Xi_e\cdot d\mu_{bos}
\label{assum}
\ee
If this assumption is true, one can indeed call $d\mu_e$
from (\ref{dmuvsBos}) the {\it NSR measure} in the
first-quantized superstring theory -- to be further
compared with the Green-Schwarz measure \cite{GS}.

In this framework the celebrated non-renormalization conjectures
\cite{Mart} imply that
\be
\sum_e A_e\Xi_e = 0
\label{vaco}
\ee
for \vspace{-0.2cm}
\be
A_e = 1
\label{A1}
\ee
or given by two combinations of Szego kernels \cite{RScalc}
$\Psi_e(x,y) =
\frac{\theta_e(\vec x-\vec y)}{\theta_e(\vec 0)E(x,y)}$
\be
A_e(x,y) = [\Psi_e(x,y)]^2
\label{Axy}
\ee
and
\be
A_e(x,y,z) = \Psi_e(x,y)\Psi_e(y,z)\Psi_e(z,x)
\label{Axyz}
\ee
For $A_e=1$ eq.(\ref{vaco}) implies the vanishing of
string-loop corrections to cosmological constant,
while (\ref{Axy}) and (\ref{Axyz}) are needed for vanishing
of string-loop corrections to 2-point and 3-point functions
respectively.
Under assumption (\ref{assum}) corrections to 1-point
function vanish automatically (without this assumption
they get contribution from correlators of $A$ with the
supercurrents).

\bigskip

{\bf 3. The case of hyperelliptic surfaces.}
If restricted to hyperelliptic locus
${\cal H}_g \subset {\cal M}_g$ of codimension $g-2$
in the moduli space ${\cal M}_g$, both Szego kernels
and DHP/CDG/G measure become proportional to rational
functions of ramification points $a_i$, $i=1,\ldots,2g+2$,
moreover all the dependence on the fermion boundary
conditions (half-integer theta-characteristic) $e$
is contained in these rational factors.
Non-vanishing contributions come only from non-singular
even characteristics, which are in one-to-one correspondence
with divisions of $2g+2$ ramification points into two
equal-size subsets
$\{a\} = \{\tilde a\}\cup\{\widetilde{\tilde a}\}$
each containing $g+1$ points.
Obviously, there are $C^{g+1}_{2g+2}$ non-singular
among the $2^{g-1}(2^g+1)$ even characteristics.
The theta-constant and
Szego kernel on hyperelliptic locus are given by
\be
\theta_e^4(0) = \det\sigma^2
\prod_{i<j}^{g+1} (\tilde a_i-\tilde a_j)
(\widetilde{\tilde a}_i - \widetilde{\tilde a}_j)
\label{Thom}
\ee
and
\be
\Psi_e(x,y) = \frac{1}{2}\left(\sqrt{\frac{u_e(x)}{u_e(y)}} +
\sqrt{\frac{u_e(y)}{u_e(x)}}\right)
\frac{\sqrt{dx dy}}{x-y}
\ee
where
\be
u_e(x) = \sqrt{\prod_{i=1}^{g+1}\frac{x-\tilde a_i}
{x-\widetilde{\tilde a}_i}}\ =\
\frac{\prod_{i=1}^{g+1} (x-\tilde a_i)}{s(x)}
\ee
and
\be
s^2(x) = \prod_{I=1}^{2g+2} (x-a_I) =
\prod_{i=1}^{g+1}(x-\tilde a_i)(x-\widetilde{\tilde a}_i)
\ee
This means that on the hyperelliptic locus ${\cal H}_g$
the vanishing relations (\ref{vaco}) for (\ref{Axy}) and (\ref{Axyz})
are reduced to
\be
\sum_e \Xi_e = 0
\label{vaco1}
\ee
and
\be
\sum_e \left(\frac{u_e(x)}{u_e(y)} + \frac{u_e(y)}{u_e(x)}\right)
\Xi_e = 0
\label{vaco2}
\ee
In what follows we often omit the transcendental factors $\det\sigma$
inessential for our considerations.

\bigskip

{\bf 4. Genus one.}
Projective transformations allow to take for
the four ramification points $\{0,1,\infty,\lambda\}$
with modular transformations restricted to six equivalences
$\lambda \sim 1-\lambda \sim 1/\lambda \sim 1/(1-\lambda)
\sim (\lambda-1)/\lambda \sim \lambda/(\lambda-1)$.
The three even characteristics, all non-singular, are
associated with the three pairs
$(0,\lambda)$, $(1,\lambda)$ and $(0,1)$. The corresponding
three theta-constants $\theta_e^4$ are
$\lambda$, $\lambda-1$ and $1$, linearly related by a single
Riemann identity
${\sum_e^3<e,*>\theta_e^4} \sim \lambda - (\lambda-1) -1 = 0$.
The DHP/CDG/G forms $\ \Xi_e = \sum_e \theta_e^{16} -
\frac{1}{2}\theta_e^8\left(\sum_e \theta_e^8\right)$
are given by $\lambda^4-\lambda^2(\lambda^2-\lambda+1) =
\lambda^2(\lambda-1)$, $-\lambda(\lambda-1)^2$,
$-\lambda(\lambda-1)$,
and (\ref{vaco1}) is obviously true:
\be
\sum_e^3 \Xi_e = \lambda(\lambda-1)\Big(\lambda-
(\lambda-1)-1\Big) = 0
\ee
At genus one the same forms $\Xi_e$ are also given by
$<e,*>\!\theta_e^4\eta^{12}$, see eq.(28) in \cite{SuMe},
and (\ref{vaco1}) is a direct consequence of Riemann identity.
As to eq.(\ref{vaco2}), it looks like\footnote{Note that
the sum in first line in (\ref{vaco2gen1}) is not vanishing,
contrary to erroneous statement in \cite{AM2} (repeated afterwards
in some subsequent papers). In fact it was the second line
of (\ref{vaco2gen1}) which was meant, used and derived there:
as rightly explained in \cite{AM2},
the sum over characteristics is equivalent to antisymmetrization
over ramification points, -- but this is true for the second,
not for the first line in (\ref{vaco2gen1}).}
$$
\sum_e^3  \frac{u_e(x)}{u_e(y)}\,\Xi_e\ \sim\
\lambda(\lambda-1)\frac{s(y)}{s(x)}\left(
\lambda\frac{x(x-\lambda)}{y(y-\lambda)} -
(\lambda-1)^4\frac{(x-1)(x-\lambda)}{(y-1)(y-\lambda)}
- \frac{x(x-1)}{y(y-1)} \right)
= \frac{\lambda^2(\lambda-1)^2(x-y)}{s(x)s(y)},
$$ \vspace{-0.5cm}
\be
\sum_e^3  \left(\frac{u_e(x)}{u_e(y)}
+ \frac{u_e(y)}{u_e(x)}\right) \Xi_e\ =\ 0
\label{vaco2gen1}
\ee

\bigskip

{\bf 5. Genus two.}
From Thomae formula (\ref{Thom}) and explicit form of
\be
\Xi_e \ \ \stackrel{\cite{DHP1,CDG}}{\sim}\ \
\frac{2}{3}\theta_e^{16} - \frac{1}{2}\theta_e^8
\sum_{e_1}^{10} \theta_{e+e_1}^8 +
\frac{1}{12}\theta_e^4 \sum_{e_1,e_2}^{10}
\theta_{e+e_1}^4\theta_{e+e_2}^4\theta_{e+e_1+e_2}^4
\ \ \stackrel{\cite{Gru}}{\sim}\ \
G^{(0)}_e - G^{(1)}_e + \frac{1}{3}G^{(2)}_e
\ee
(see \cite{SuMe} for  normalization conventions and other
details) we deduce:
\be
\!\!\!\! {\bf g=2:}\ \ \ \
\Xi_e \sim \frac{1}{2}\,\theta^4_e(0)
\left( \sum_{i\neq j\neq k}^{g+1} \tilde a_i\tilde a_j\tilde a_k
- \sum_{i\neq j}^{g+1}\tilde a_i\tilde a_j
\sum_k^{g+1}\widetilde{\tilde a}_k
+ \sum_i^{g+1} \tilde a_i \sum_{j\neq k}^{g+1}
\widetilde{\tilde a}_j\widetilde{\tilde a}_k -
\sum_{i\neq j\neq k}^{g+1}
\widetilde{\tilde a}_i\widetilde{\tilde a}_j\widetilde{\tilde a}_k
\right)
\prod_{I<J}^{2g+2} (a_I-a_J)
\label{factorXigen2}
\ee
\vspace{-0.3cm}
$$
\theta^4_e(0) \sim \prod_{i<j}^{g+1}(\tilde a_i-\tilde a_j)
(\widetilde{\tilde a}_i-\widetilde{\tilde a}_j)
$$
For example,
$$
\Xi_{123|456}\ \sim \
(a_{12}a_{13}a_{23})^2(a_{45}a_{46}a_{56})^2
\Big(3a_1a_2a_3 - (a_1a_2+a_1a_3+a_2a_3)(a_4+a_5+a_6) +
$$ $$
+ (a_1+a_2+a_3)(a_4a_5+a_4a_6+a_5a_6) - 3a_4a_5a_6\Big)
\cdot a_{14}a_{15}a_{16}a_{24}a_{25}a_{26}a_{34}a_{35}a_{36}
$$
Note that this expression is symmetric under permutation of
two triples $123$ and $456$ and also under permutations of any
two points inside the triple: in this sense it is indeed
a function of a given characteristic.
We now apply the argumentation of \cite{AM2} to prove
the vanishing theorems (\ref{vaco1}) and (\ref{vaco2}).
Sum over $e$ is symmetrization w.r.t. {\it all} permutations
of all the six ramification points:
\be
\sum_e \Xi_e \ \ \sim \sum_{{\rm perms\ of}\ abcdef}
\Xi_{abc|def}
\ee
Since all $\Xi_e$ have a common factor
$\Pi(a)\equiv \prod_{I<J}^{2g+2}(a_I-a_J)$,
which is {\it anti}symmetric w.r.t. the two points permutation,
$\Pi(a)^{-1} \sum_e \Xi_e$ should be an
{\it anti}symmetric function, i.e. should itself be proportional
to $\Pi(a)$. However, this sum has degree
$9$ in $a_I$, while the $\Pi(a)$ has degree $15$ --
and this means that
\be
{\bf g=2:}\ \ \ \ \ \sum_e^{10} \Xi_e = 0
\ee
Similarly,
\be
\sum_e \left(\frac{u_e(x)}{u_e(y)} + \frac{u_e(y)}{u_e(x)}\right)
\Xi_e = \frac{1}{s(x)s(y)}\sum_e \left(
\prod_{i=1}^{g+1}(x-\tilde a_i)(y-\widetilde{\tilde a}_i)
+ \prod_{i=1}^{g+1}(x-\widetilde{\tilde a}_i)(y-\tilde a_i)
\right)\Xi_e
\label{uuXi}
\ee
should be proportional to $(x-y)\Pi(a)^2$, since it should
vanish at $x=y$, what is impossible because the sum has degree
$15$ in $x,y,a_I$, what is less than $16$ -- the degree of
$(x-y)\Pi(a)^2$. This means that the sum (\ref{uuXi}) vanishes,
and (\ref{vaco2}) is true for $g=2$.
Note that for this standard argument \cite{AM2} to work
it is important that $\Pi(a)$
is factored out from (\ref{factorXigen2}):
if this did not happen, such simple calculus would not work.
At the same time, the factorization of $\theta_e^4$ is not
important: even if there was no such factor, the argument
would perfectly work.

\bigskip

{\bf 6. From genus $g$ to genus $g-1$,
e.g. from genus two to genus one.}
Degeneration of hyperelliptic surfaces and
factorization of theta-constants in Thomae formula
look a little tricky: this is the price to be paid for
simplifications of Riemann and other relations between
theta-constants in this parametrization.
The simple degeneration is ${\cal H}_g \rightarrow {\cal H}_{g-1}$
-- shrinking of a handle: then just some two of ramification
points approach each other and the corresponding cut turns into
a puncture. However, from the point of view of string measures
this degeneration is not the simplest one, because it is associated
with insertion of non-trivial vertex operators, describing
propagation of string excitations with non-trivial spins
along the shrinking handle.
Much simpler from this point of view is degeneration
${\cal H}_{g_1+g_2} \rightarrow {\cal H}_{g_1}\times {\cal H}_{g_2}$,
when no spin-carrying particles can propagate along the long tube
and measures simply factorize, in particular,
$\Xi_e^{(g_1+g_2)} \rightarrow \Xi_{e_1}^{(g_1)}\Xi_{e_2}^{(g_2)}
+ \ldots $ However, the corresponding behavior of ramification
points is more involved. In what follows we consider only the
simplest degeneration of this kind, ${\cal H}_g \,\rightarrow
{\cal H}_1\times{\cal H}_{g-1}$, all others can be analyzed in the
same way.

In this limit some {\it three} ramification points, let them be
$A_1=a_{2g}$, $A_2=a_{2g+1}$ and $A_3=a_{2g+2}$ approach each other
and finally become a single point $A$ (they are sometimes underlined
in formulas below).
The $2g$ ramification points of emerging curve of genus $g-1$
are $a_1,\ldots,a_{2g-1},A$, while the decoupling torus
(genus-one curve) is characterized by the set $A_1,A_2,A_3,\infty$,
so that $A_{23} = \lambda A_{13}$, $A_{12} =
(1-\lambda)A_{13}$, and $A_{12} \rightarrow 0$, while
$\lambda$ remains finite and becomes the modulus of the torus.
For $g=2$ the $10$ theta-constants behave in this limit as follows:
\be
\theta^4_{123|\underline{456}}
= {\det}^2\sigma\ a_{123}a_{456} = {\det}^2\sigma\
a_{123} A_{123} = \lambda(1-\lambda) a_{123}
\cdot\Big(A_{13}^3 {\det}^2\sigma\Big)
\longrightarrow 0
\ee
and, say,
$$
\theta^4_{12\underline{4}|3\underline{56}}
= {\det}^2\sigma \ a_{124}a_{356} \rightarrow
(1-\lambda) a_{12} (A-a_1)(A-a_2)(A-a_3)^2
\cdot \Big(A_{13}{\det}^2\sigma\Big)
\Big(1 + O(A_{13})\Big) \longrightarrow 
$$ \vspace{-0.6cm}
\be
\longrightarrow
\underbrace{(1-\lambda)}_{\theta^{(1)}_{1\lambda|0\infty}}
\underbrace{a_{12}(A-a_3)}_{\theta^{(1)}_{12|3A}}
\underbrace{(A-a_1)(A-a_2)(A-a_3)}_{P(A)}P(A)^{-1}\ \sim\
\theta^{(1)}_{1\lambda|0\infty}\theta^{(1)}_{12|3A}
\ee
and similarly for the other $8$ characteristics where
the three underlined points appear among both $\tilde a$
and $\widetilde{\tilde a}$.
In taking these limits we used that
\be
\det\sigma^{(g)} \ \stackrel{A_{13}\rightarrow 0}\longrightarrow\
\frac{1}{\sqrt{A_{13}P(A)}}\det\sigma^{(1)}\det\sigma^{(g-1)}
\label{detsiglim}
\ee
To understand this behavior one needs to recall that
$\sigma$ is a matrix of $A$-periods of $g$ non-canonical
holomorphic differentials $\frac{x^{0,1,\ldots,g-1}dx}{s(x)}$.
In the limit of our interest $s^{(g)}(x) \rightarrow
(x-A)s^{(g-1)}(x)$ and only $g-1$ of these $g$ differentials,
namely
\be
\frac{(x-A)x^{0,1,\ldots,g-2}dx}{s^{(g)}(x)}\ \longrightarrow\
\frac{x^{0,1,\ldots,g-2}dx}{s^{(g-1)}(x)}
\ee
behave smoothly and turn into the same kind of holomorphic
differentials on the genus $g-1$ curve.
The last differential naively looks singular,
$\frac{dx}{s(x)} \rightarrow \frac{dx}{(x-A)s^{(g-1)}(x)}$,
however the singularity is actually resolved while the three
points $A_1,A_2,A_3$ remain different.
In appropriate coordinate $x = A + A_{13}\chi$ this differential
looks like the ordinary holomorphic differential on the torus,
$\frac{1}{\sqrt{A_{13}P(A)}}
\frac{d\chi}{\sqrt{\chi(\chi-1)(\chi-\lambda)}}$ only with
the coefficient,  which diverges as $A_{13}\rightarrow 0$.
Therefore the matrix $\sigma^{(g)}$ consists of non-singular
$\sigma^{(g-1)}$ and an extra singular element
$\frac{\sigma^{(1)}}{\sqrt{A_{13}P(A)}}$ on diagonal.
Non-singular elements in associated row and column do not
affect the singular asymptotics (\ref{detsiglim}) of $\det \sigma$.

It remains to see what happens to $\rho_e$, and here we again
return to genus two. Since $\theta_{123|\underline{456}}$ vanishes
in the limit, the behavior of $\rho_{123|\underline{456}}$
is of no interest.
Thus we need to look only at
$$
\rho_{12\underline{4}|3\underline{56}} =
3a_1a_2\underline{a_4}
- (a_1a_2 + a_1\underline{a_4} + a_2\underline{a_4})
(a_3+\underline{a_5} + \underline{a_6}) +
(a_1+a_2+\underline{a_4})(a_3\underline{a_5} + a_3\underline{a_6}
+\underline{a_5a_6}) - 3a_3\underline{a_5a_6} \longrightarrow
$$
$$
\longrightarrow\ 3(Aa_1a_2-A^2a_3)
- \big(a_1a_2+A(a_1+a_2)\big)(a_3+2A) +
(a_1+a_2+A)(2Aa_3 + A^2) =
$$ \vspace{-0.6cm}
\be
= (A-a_1)(A-a_2)(A-a_3) = P(A)
\label{factgen2}
\ee

\bigskip

{\bf 7. Other genera.}
In general one can {\it assume} that on hyperelliptic locus
\be
\Xi_e = \ \det\sigma^8
R_e \prod_{I<J}^{2g+2} (a_I-a_J)
\label{XiR}
\ee
Since by Thomae formula the weight-two form $\theta_e^4$ has
degree  $g(g+1)$ in ramification points, the weight-eight
form $\Xi_e$ has degree $4g(g+1)$,
while the product $\prod_{i<j}^{2g+2} a_{ij}$
has degree $(g+1)(2g+1)$.
Thus $R_e$ is a polynomial of degree $g(g+1)+g^2-1=(g+1)(2g-1)$
in ramification points.
Explicit expression for it can be straightforwardly deduced
from the theta-constant expression (\ref{dmuGru}),
but this exercise remains beyond the scope of the present
paper: it is a very important problem for the future research.
The most important question is if the $\Pi(a)$
factor is indeed extracted from $\Xi_e$ on the
hyperelliptic locus. If this happens-- as we now {\it assume} --
then the simple
power-counting of \cite{AM2}, reminded above for the simples
case of $g=2$, is sufficient to prove the two vanishing identities
(\ref{vaco1}) and (\ref{vaco2}). Indeed, we have two
polynomials
\be
\sum_e \Xi_e \ \stackrel{(\ref{XiR})}{\sim}\ \sum_e R_e
\ = 0
\ee
and
\be
s(x)s(y)\sum_e \left(\frac{u_e(x)}{u_e(y)} +
\frac{u_e(y)}{u_e(x)}\right)\Xi_e
\ \stackrel{(\ref{XiR})}{\sim}\  \sum_e R_e \left(
\prod_{i=1}^{g+1}(x-\tilde a_i)(y-\widetilde{\tilde a}_i)
+ \prod_{i=1}^{g+1}(y-\widetilde{\tilde a}_i)(x-\tilde a_i)\right)
\ = 0
\ee
of degrees  $(g+1)(2g-1)$ and $(g+1)(2g+1)$ respectively,
and they should be {\it anti}symmetric under permutations of
all ramification points, i.e. should be divisible by $\Pi(a)$
which has degree $(g+1)(2g+1)$. It follows, that the first
polynomial vanishes and, as a corollary, the second one should
be additionally divisible by $(x-y)$ -- and then it also needs
to vanish.

At genus $g=2$ the polynomial $R_e$ is further decomposed
as $R_e = \theta_e^4 \rho_e$, but this need not happen for
higher genera. We emphasize once again that this does not affect
the validity of pre-renorminvariance theorems (\ref{vaco1})
and (\ref{vaco2}). This observation, though confined to
the hyperelliptic locus, can still be an additional source of
optimism about the DHP-CDG-G  conjecture \cite{DHP1,CDG,Gru}:
even without the $\theta^4_e$ factors {\it some} important
properties of the correlators can survive.

\bigskip

{\bf 8. A little more on $g \rightarrow 1+(g-1)$
factorization of hyperelliptic measures.}
Of $C^{g+1}_{2g+2}$ theta-constants with non-singular characteristics
the $C^{g-2}_{2g-1}$ of the form
$\theta_{123\ldots|\ldots}$ vanish in degeneration limit, while
the remaining $3C^{g-1}_{2g-1}$ turn into
$$
\theta_{A_1A_2\underbrace{\ldots}_{g-1}|
A_3\underbrace{\ldots}_g}^{(g)}\ \longrightarrow \ 
{\det}^2\sigma \ A_{12} \prod_{i<j}^{g-1}{\tilde a}_{ij}
\prod_{i<j}^g \widetilde{\tilde a}_{ij}
\prod_{i=1}^{g-1} (A-\tilde a_i)^2
\prod_{i=1}^g (A-\widetilde{\tilde a}_i)
= $$ \vspace{-0.6cm}
\be
=\underbrace{{\det}^2\sigma \ A_{12}}_{
\theta^{(1)}_{12|3\infty}/P(A)} 
\underbrace{
\left(\prod_{i=1}^{g-1} (A-\tilde a_i)\prod_{i<j}^{g-1}
\tilde a_{ij}\right)
\prod_{i<j}^g \widetilde{\tilde a}_{ij}}_{\theta^{(g-1)}}
\underbrace{ \prod_{I=1}^{2g-1}(A-a_I)}_{P(A)}
\ \sim\ \theta^{(1)}_{12|3\infty}
\theta^{(g-1)}_{A\underbrace{\ldots}_{g-1}|
\underbrace{\ldots}_g} 
\ee
where $P(A) \equiv \prod_{I=1}^{2g-1}(A-a_I)$
is a specific auxiliary polynomial of degree $2g-1$.
It follows that
\be
\Xi_{e_g}^{(g)} \ \longrightarrow\ 
\Xi_{e_1}^{(1)}\Xi_{e_{g-1}}^{(g-1)}
\label{Xifac}
\ee
At the same time
\be
\Pi^{(g)}(a) \longrightarrow
\lambda(1-\lambda) A_{13}^3 \prod_{I<J}^{2g-1}
(a_I-a_J) \prod_{I=1}^{2g-1} (A-a_I)^3 = A_{13}^3\cdot
\Pi^{(1)}\Pi^{(g-1)}P(A)^2
\label{Pifac}
\ee
(one $P(A)$ is absorbed into $\Pi^{(g-1)}$). 

From (\ref{Xifac}), (\ref{Pifac}) and (\ref{detsiglim})
it follows that $R_e$ in (\ref{XiR}) should factorize as
\be
R_{e_g}^{(g)} \longrightarrow
R_{e_1}^{(1)} R_{e_{g-1}}^{(g-1)} P^2(A) A_{13}
\label{factorg}
\ee
where the genus one $R_e^{(1)} = \theta_e^4/{\det}^2\sigma$ 
(i.e. $\rho_e^{(1)}=1$).
In comparison to (\ref{factgen2}) one should remember that
at genus two $R_e{\det}^2\sigma = \theta_e^4 \rho_e$, the
$A_{13}$ factor is absorbed into factorization of $\det\sigma$,
and $\rho_e$ transforms with a single $P(A)$ factor.
Factorization (\ref{factorg}) respects the power counting:
$$
(g+1)(2g-1) = g(2g-3) + 2(2g-1) + 1
$$
(if $\rho_e$ could exist, i.e. if $\theta_e^4$ factors out from
$\Xi_e$ and $R_e$ at least on hyperelliptic locus, then it would
be a polynomial of degree $g^2-1=(g-1)(g+1)$ in ramification
points and factorization $g\rightarrow 1+(g-1)$ would imply the
following power counting: $g^2-1 = (g-1)^2-1 + 2g-1$.)

\bigskip

{\bf 9. Conclusion.} To summarize, we considered restriction
of DHP-CDG-G NSR measures on hyperelliptic locus and argued that
validity of identities (\ref{vaco1}) and (\ref{vaco2}), underlying
the non-renormalization theorems is not too much affected by the
violation of the "$\theta_e^4$-hypothesis" by these measures for
$g\geq 3$. The argument is not a full proof because it uses a
(plausible) assumption that $\Pi(a)$ is factored out from the
hyperelliptic measure (\ref{XiR}). This assumption can be
checked by substitution of Thomae formula
(\ref{Thom}) into (\ref{dmuGru}), what is straightforward but
somewhat tedious exercise (in this paper it is performed only
for $g=2$).
Alternatively information about the structure of $R_e$ can be
provided by explicit generalization of factorization relation
(\ref{factgen2}) to (\ref{factorg}) for $g>2$.
These exercises deserve to be done in any case,
also because the explicit knowledge of $R_e$
(presumably, a nice expression) will be of direct use for evaluation
of $4$- and higher-point functions, at least on the hyperelliptic
locus. One of the main goals of this letter is to remind about
the power and simplicity of hyperelliptic calculus and advocate
its application in the new attack on the bastions of the
first-quantized string theory.

\section*{Acknowledgements}

This work  is partly supported by Russian Federal Nuclear
Energy Agency and Russian Academy of Sciences,
by the joint grant 06-01-92059-CE,  by NWO project 047.011.2004.026,
by INTAS grant 05-1000008-7865, by ANR-05-BLAN-0029-01 project,
by RFBR grant 07-02-00645 and
by the Russian President's Grant of Support for the Scientific
Schools NSh-3035.2008.2

\end{document}